\documentclass[a4paper]{jpconf}  % for double-spaced preprint
\usepackage[dvips]{graphicx}
\usepackage{dcolumn}   % needed for some tables
\usepackage{bm}        % for math
\usepackage{amssymb}   % for math
\usepackage[utf8]{inputenc}
\usepackage{mathrsfs}
\usepackage{amsmath}
\usepackage{hyperref}
\usepackage{textcomp}
\usepackage{subfigure}
\usepackage{color}
\usepackage{cite}

\graphicspath{{pictures/}}

\begin{document}

\title{Transitions between topologically non-trivial configurations}
\author{V A Gani,$^{1,2}$ A A Kirillov$^{1,3}$ and S G Rubin$^{1,4}$ }
\address{$^1$National Research Nuclear University MEPhI (Moscow Engineering Physics Institute), 115409 Moscow, Russia}
\address{$^3$National Research Center Kurchatov Institute, Institute for Theoretical and Experimental Physics, 117218 Moscow, Russia}
\address{$^3$Department of Theoretical Physics, Yaroslavl State P.G. Demidov University,\\ 150003 Yaroslavl, Russia}
\address{N.I. Lobachevsky Institute of Mathematics and Mechanics, Kazan Federal University,\\ 420008 Kazan, Russia}
\ead{vagani@mephi.ru}

\begin{abstract}
We study formation and evolution of solitons within a model with two real scalar fields with the potential having a saddle point. The set of these configurations can be split into disjoint equivalence classes. We give a simple expression for the winding number of an arbitrary closed loop in the field space and discuss the evolution scenarios that change the winding number. These non-trivial field configurations lead to formation of the domain walls in the three-dimensional physical space.
\end{abstract}

\section{Introduction and Motivation}

Recently, the inflationary models with several fields have become the subject of a growing interest \cite{Abbyazov}. Many of these models involves complicated potentials, which arise in various supersymmetric models \cite{Kawai}, in the brane approach \cite{Liddle0,Lust}, and in the landscape paradigm \cite{susskind01,Vil}. The inflation within the multi-field models has been thoroughly elaborated, its effect on fluctuations of the cosmic microwave background was analyzed \cite{Liddle}. One of the important features of the multi-field inflationary models is a potential with a number of minima \cite{Liu.2015}. A field-theoretical system with a potential with many minima can possess topologically non-trivial scalar field configurations --- solitons \cite{manton}. Such configurations could form domains with extra energy density or collapse into black holes after the end of the inflation. It is important that non-trivial field configurations could be formed even if the potential of the model has only one minimum. In this paper we study the conditions at which these solitons can be formed.

In this paper we consider a system of two real scalar fields, $\varphi$ and $\chi$. As we show, the domain wall can be formed after the inflation even in the case of only one potential minimum. It is demonstrated that non-trivial field configurations can appear if the fields tend to the same minimum of the potential at different spatial asymptotics. At the same time, the existence of a saddle point of the potential is necessary for this phenomenon to occur. We introduce the topological classification, which allows us to divide non-trivial configurations into disjoint (homotopic) classes labeled by {\it winding number}. We also find the conditions at which the winding number can change, i.e.\ the configuration can move from one homotopic class to another. It is important to note that such transitions are not possible if the fields tend to different potential minima at different spatial asymptotics. Only transitions due to quantum tunneling are possible in this case \cite{Dine}.

The fields $\varphi$ and $\chi$ live in the $(3+1)$-dimensional space-time. Nevertheless, we will study one-dimensional configurations. Thus the multi-field inflationary models are connected with $(1+1)$-dimensional field-theoretical models, where many important results have been obtained recently, in particular, related to soliton interactions \cite{GaKuPRE,GaKuLi,Aliakbar,GaLeLi,GaLeLiconf,Belendryasova.arXiv.08.2017,Bazeia.arXiv.2017.sinh,Gani.arXiv.2017.dsg,Radomskiy}, soliton stability \cite{kurochkin}, bubble decay \cite{Gani.YaF.1999,Gani.YaF.2001}, and planar domain walls \cite{GaKu.SuSy.2001,GaKsKu01,GaKsKu02,lensky,GaLiRa,GaLiRaconf}, see also book \cite{manton} and review \cite{aek01}.

\section{The Model}

Consider a model with two real scalar fields $\varphi$, $\chi$ in $(3+1)$ space-time dimensions, with its dynamics determined by the Lagrangian density
\begin{equation}
	\label{eq:Lagrangian}
	\mathcal{L} = \frac{1}{2} g^{\mu\nu} \left(  \partial_{\mu}\varphi\partial_{\nu}\varphi + \partial_{\mu}\chi\partial_{\nu}\chi \right)
    	- V(\varphi,\chi),
     \end{equation}
with the Minkowski metric tensor. The equations of motion following from the Lagrangian \eqref{eq:Lagrangian} are
\begin{equation}
	\begin{cases}
		\varphi_{tt} + 3 H \varphi_t - \varphi_{rr} 
        	-\cfrac{2}{r}\:\varphi_r 
            = -\cfrac{\partial V}{\partial\varphi},\\
		\chi_{tt} + 3 H \chi_t - \,\chi_{rr} 
        	-\cfrac{2}{r}\:\chi_r 
            = - \cfrac{\partial V}{\partial\chi},
	\end{cases}
    \label{eq:Equations_1}
\end{equation}
where $H=\dot{a}/a$ is the Hubble parameter, which is small after the end of the inflation. The friction term ensures the dumping of long-period oscillations during the reheating stage. Let we have a domain $\mathcal B$, which is surrounded by the border $\partial\mathcal B$. In this border, the fields are at the saddle point $(\varphi_\text{s},\chi_\text{s})$ of the potential $V(\varphi,\chi)$. Below we demonstrate that a non-trivial field configuration can be formed even in the case of the same vacuum state at both sides of the border, i.e.\ inside and outside of the domain $\mathcal B$.

Below we assume that the cosmological expansion is small, and we use the physical distances $R=a(t)r$. Besides that, we study configurations of the type of planar domain walls, because for an observer near the border $\partial\mathcal{B}$ the wall is almost flat. According to this, the equations \eqref{eq:Equations_1} become
\begin{equation}
	\begin{cases}
		\varphi_{tt} + 3 H \varphi_t - \varphi_{xx} = - \displaystyle\frac{\partial V}{\partial\varphi},\\
		\chi_{tt} + 3 H \chi_t - \chi_{xx} = - \displaystyle\frac{\partial V}{\partial\chi}.
	\end{cases}
    \label{eq:Equations}
\end{equation}
For our numerical simulations we use the following potential $V(\varphi,\chi)$ with a saddle point:
\begin{equation}
	\label{eq:potential}
	V(\varphi,\chi) = d\,(\varphi^2 + \chi^2) + 
    	a\:\exp{[-b\:(\varphi-\varphi_0^{})^2 - c\:(\chi-\chi_0^{})^2]},
\end{equation}
where $a$, $b$, $c$, $d$ are positive parameters of the model, and the point $(\varphi_0^{},\chi_0^{})$ is the local maximum of the potential \eqref{eq:potential}. In our simulations we used the following values of the parameters: $a = 5\cdot 10^2$, $b = 1$, $c = 1$, $d = 1$, $\varphi_0^{} = -5$, $\chi_0^{} = 0$.
\begin{figure}
\begin{center}
	\includegraphics[width=0.5\textwidth]{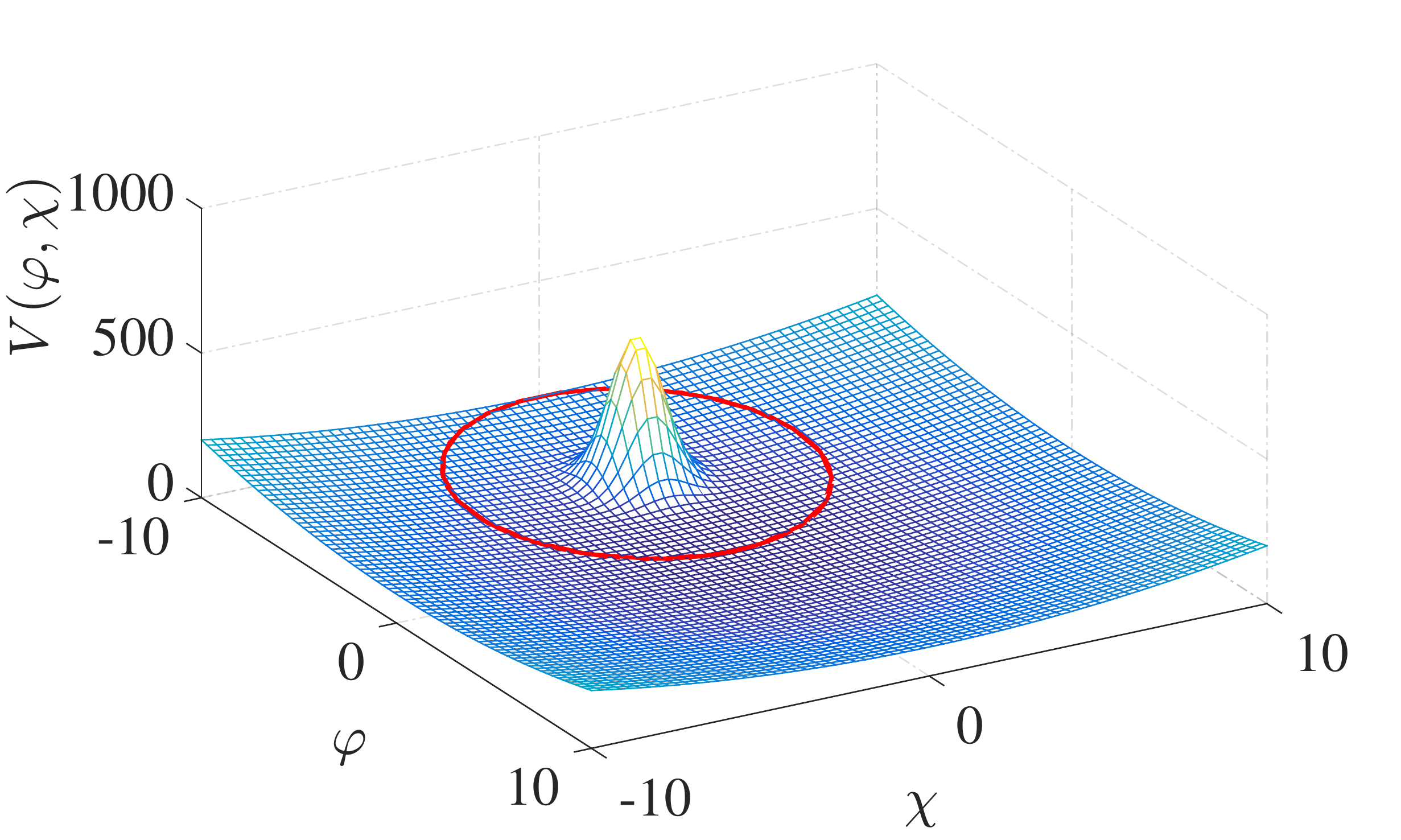}
\end{center}
\caption{The potential \eqref{eq:potential} with the initial field configuration \eqref{eq:InCond}, \eqref{eq:Theta}.}
\label{fig:fig1}
\end{figure}

\section{Topological classification and numerical simulations}

We can split the set of all closed curves in the $(\varphi,\chi)$ plane into different equivalence classes (homotopic classes). In the spirit of this splitting, the equivalent curves have the same winding (or topological) number $N$, which is the number of turns of the curve around the local maximum of the potential, i.e.\ around the point $(\varphi_0^{},\chi_0^{})$. Then we use the residue theorem in order to obtain an explicit expression for $N$. According to this theorem, the line integral of some function $f(\zeta)$ of the complex variable $\zeta=\varphi+i\chi$ around the closed curve in the $(\varphi,\chi)$ plane is equal to the sum of residues of $f(\zeta)$ at isolated singular points (surely, each counted as many times as the curve winds around the point) multiplied by $2\pi i$. We take the function $f(\zeta)=\frac{1}{\zeta-\zeta_0^{}}$. After some simple algebra, we obtain:
\begin{equation}
	N[\varphi,\chi] = \frac{1}{2\pi}\int\limits_{-\infty}^{+\infty}
    \frac{(\varphi-\varphi_0^{})
	\chi_x^{} -(\chi-\chi_0^{})
	\varphi_x^{}}{(\varphi-\varphi_0^{})^2+(\chi-\chi_0^{})^2}\:dx.
    \label{eq:NumWound}
\end{equation}
Thus we have split all closed curves in the $(\varphi,\chi)$ plane into homotopic (equivalence) classes, depending on the number of windings of the curve around the local maximum of the potential $(\varphi_0^{},\chi_0^{})$.

In our numerical simulations we start from an initial field configuration, which is a closed loop in the $(\varphi,\chi)$ plane. According to this, the boundary conditions are
\begin{equation}
	\begin{cases}
		\varphi(-\infty,t) = \varphi(+\infty,t),\\
        \varphi_x^{}(-\infty,t) = \varphi_x^{}(+\infty,t),\\
		\chi(-\infty,t) = \chi(+\infty,t),\\
        \chi_x^{}(-\infty,t) = \chi_x^{}(+\infty,t).
	\end{cases}
    \label{eq:BondCond}
\end{equation}
Consider the initial configuration, which encircles the potential maximum:
\begin{equation}
	\begin{cases}
		\varphi(x) = \varphi_0^{} + R\cos\theta(x),\\
		\chi(x) = \chi_0^{} + R\sin\theta(x),
	\end{cases}
    \label{eq:InCond}
\end{equation}
where $\theta(x)$ is
\begin{equation}
	\theta(x) = N \pi \left( 1 + \tanh x \right).
    \label{eq:Theta}
\end{equation}
Here $R>0$ is the radius of the circle, and $0\le\theta(x)\le 2\pi N$ at $-\infty\le x\le+\infty$, see fig.~\ref{fig:fig1}.

\section{Results}

\begin{figure}
\begin{center}
\includegraphics[width=52mm]{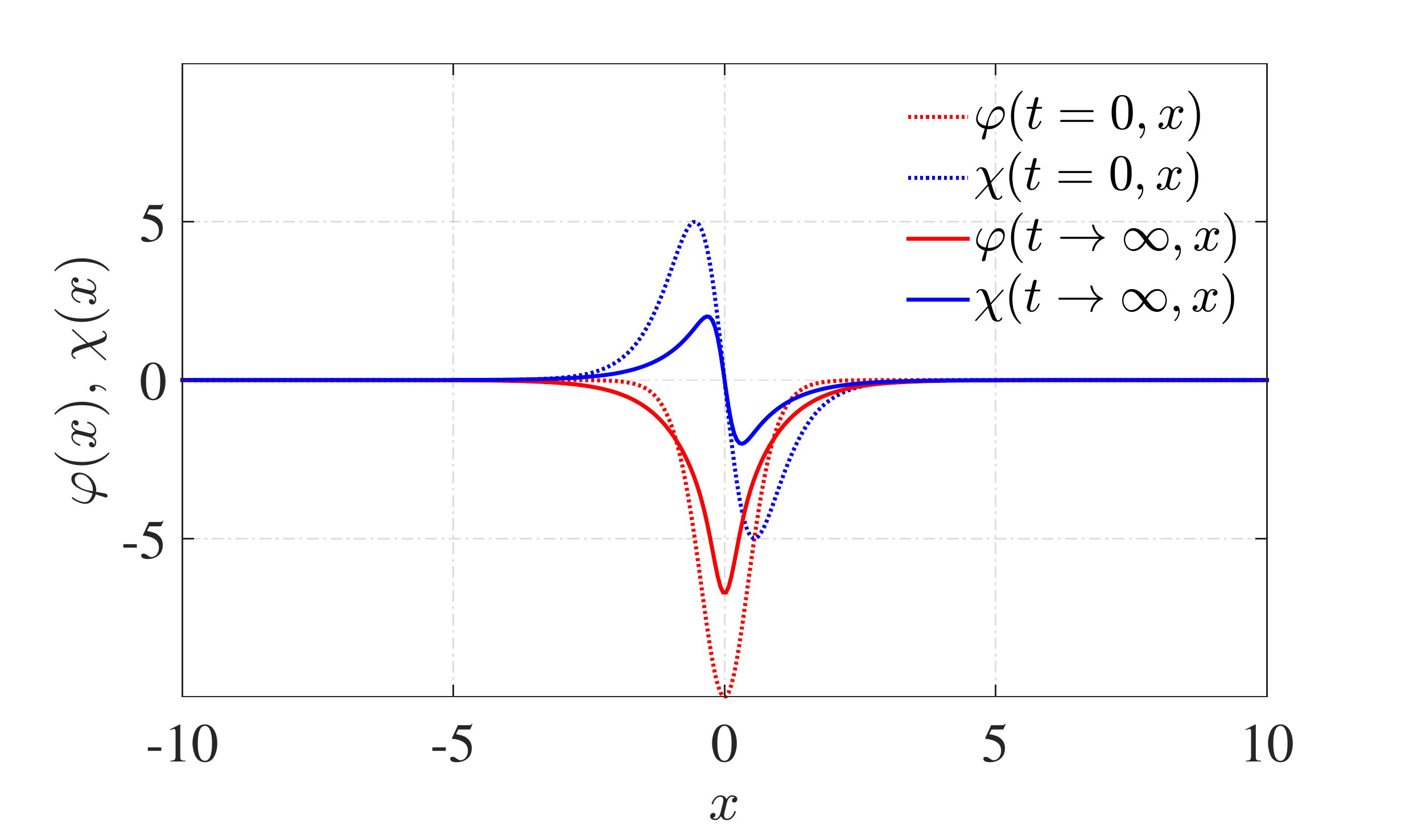}
\includegraphics[width=52mm]{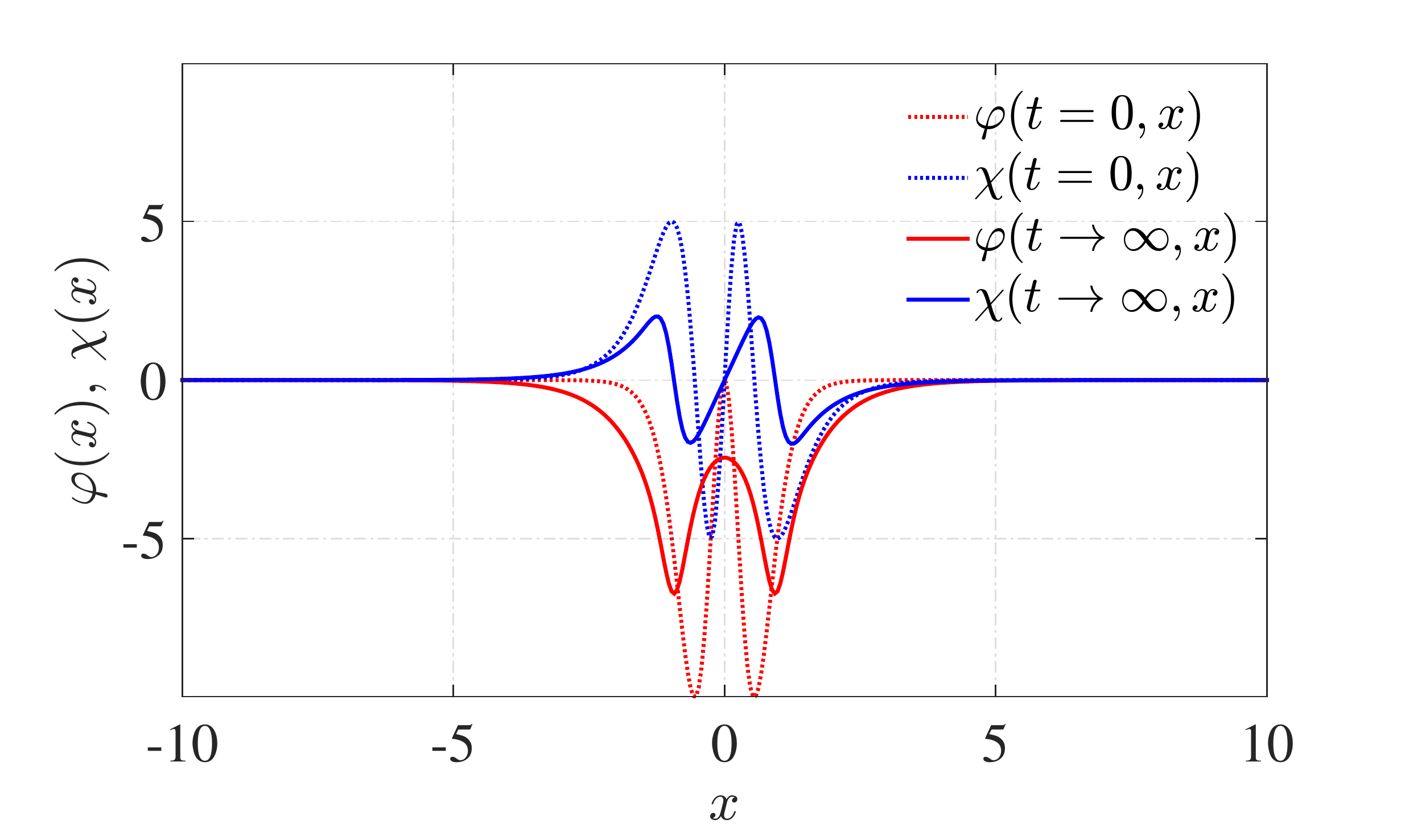}
\includegraphics[width=52mm]{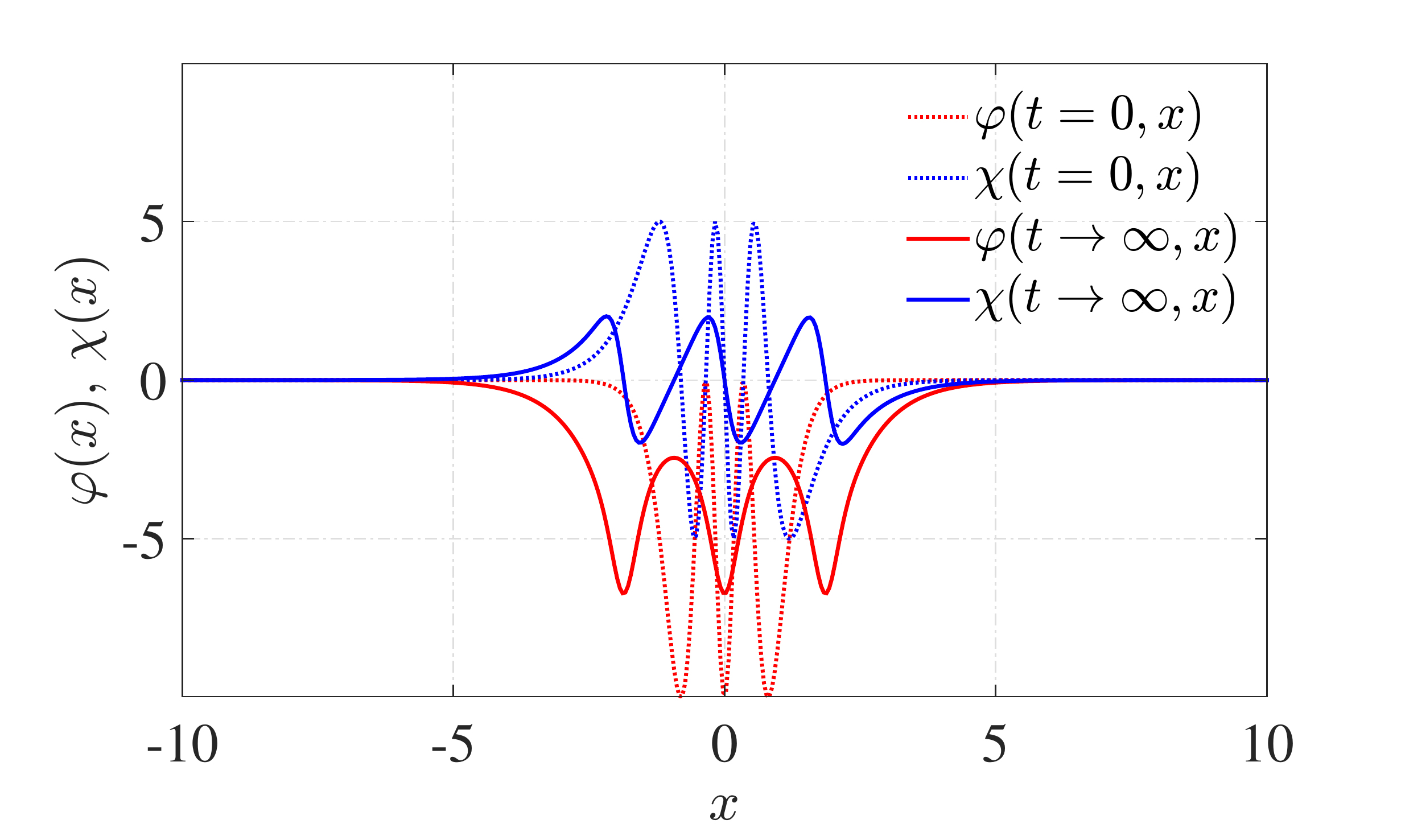}
\end{center}
\begin{center}
\includegraphics[width=52mm]{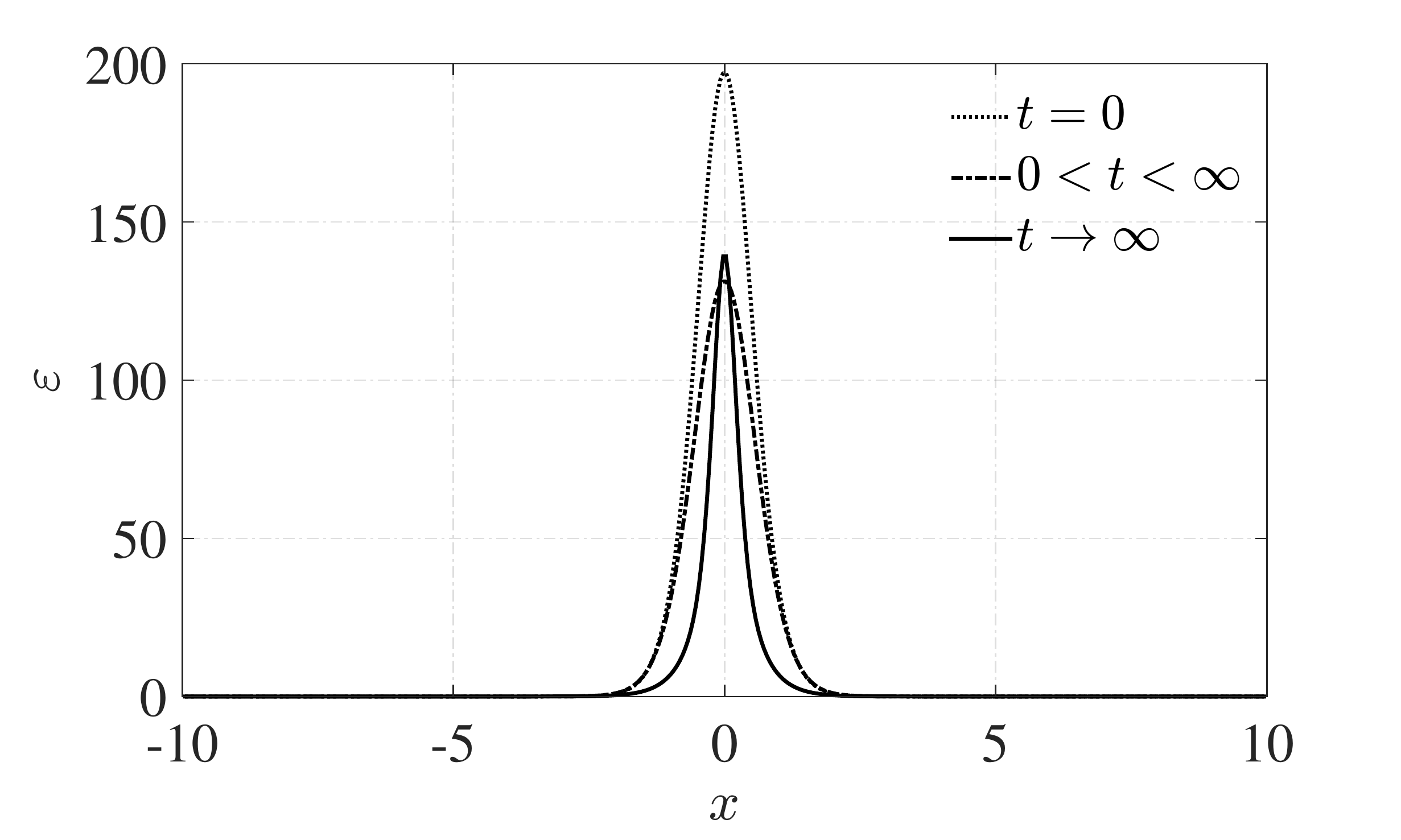}
\includegraphics[width=52mm]{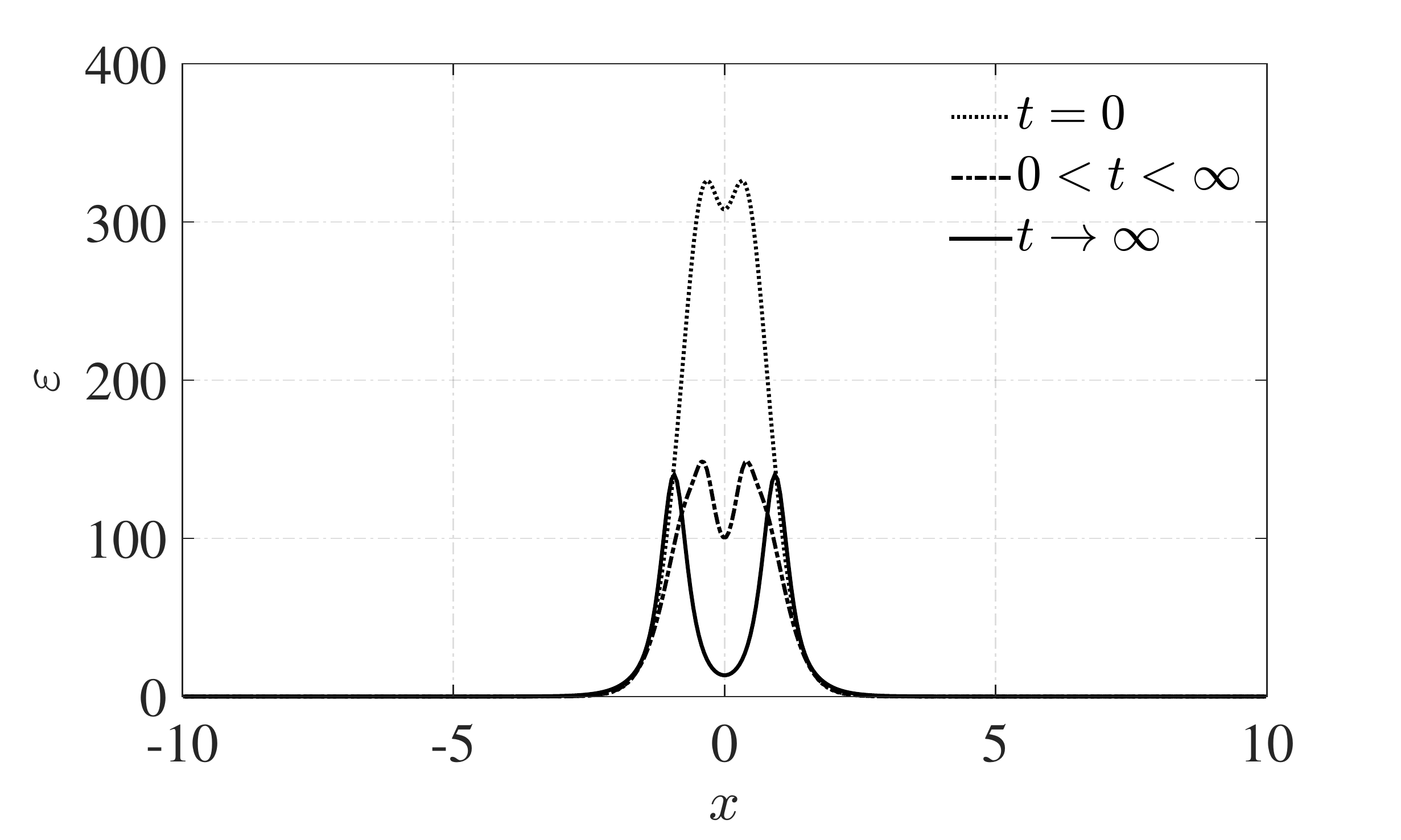}
\includegraphics[width=52mm]{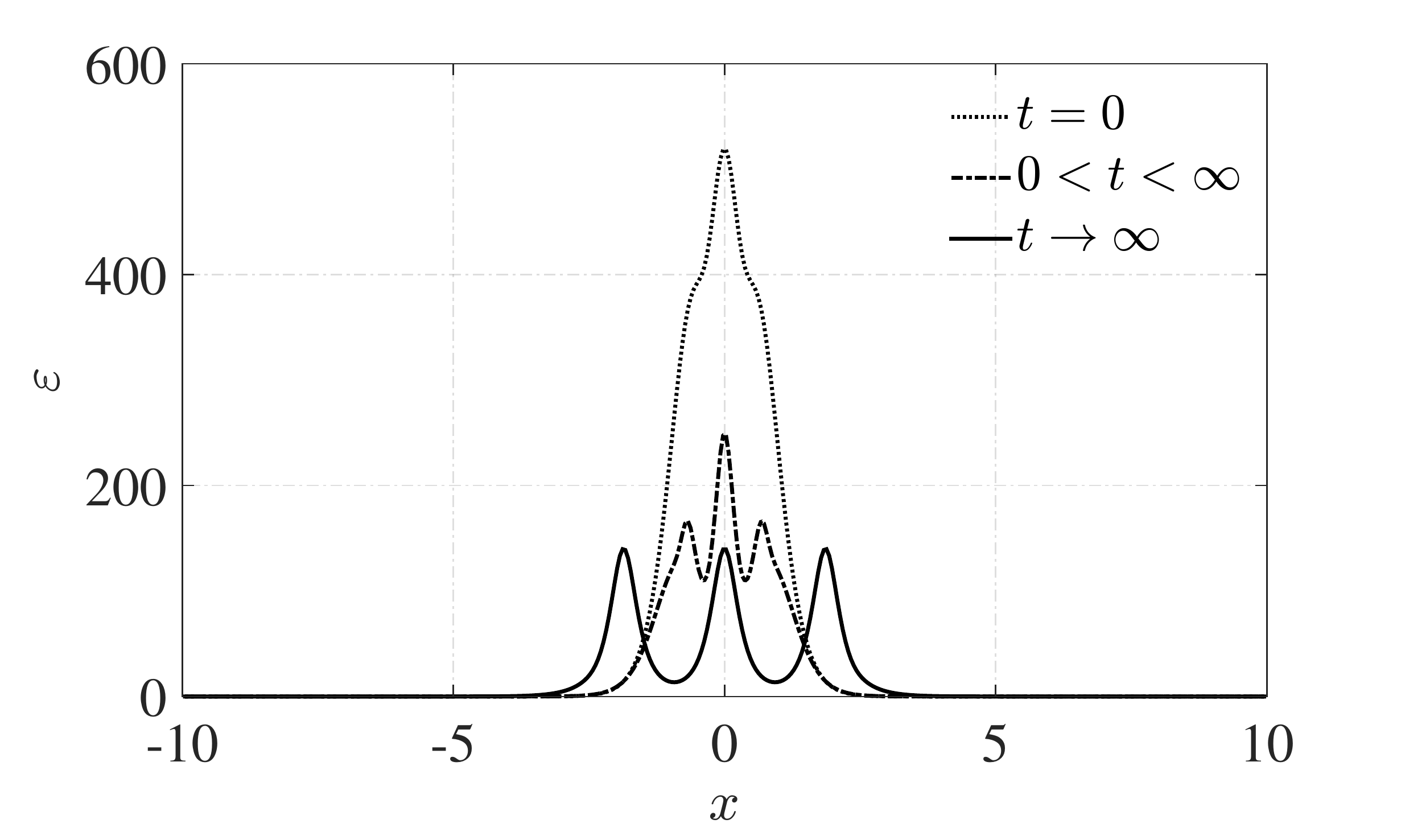}
\end{center}
\caption{Top: spatial distribution of the fields at $t=0$ and at $t\to +\infty$ for the potential \eqref{eq:potential} for various winding numbers $N=1,2,3$. Bottom: the corresponding energy density $\varepsilon$ of the solitons as a function of the spatial coordinate $x$.}
\label{fig:fig2}
\end{figure}

We studied the evolution of the initial configuration \eqref{eq:InCond}, \eqref{eq:Theta} for winding numbers $N=1,2,3$ numerically. At all these winding numbers we observed the tightening of the closed loop around the potential maximum $(\varphi_0^{},\chi_0^{})$. This result means that a space trajectory connecting two points inside and outside of the domain $\mathcal B$ necessarily goes through a domain wall or through a series of domain walls, at $N=1$ and at $N\ge 2$, respectively, see fig.~\ref{fig:fig2}.

\section{Conclusion}

A finite height of the maximum allows the change of the homotopic class of a given field configuration, along with the stepwise decrease of the winding number. The realisation of such changes depends on the parameters of the potential and on those of the initial field configuration. We have quantitatively confirmed these qualitative features of the transitions that change the winding number.

Field configurations similar to those considered in this work could be formed during the inflationary stage, and lead to strong inhomogeneities in the CMB radiation temperature fluctuations.

This new type of solitons that has been studied in this paper may be also useful as a mechanism of BHs with complex mass spectra production  \cite{RubKhlopSakh_PBHProd,KhlopRubSakh,Belotsky,PBH} and could be used for solution of a number of cosmological and astrophysical problems, e.g., high-$z$ quasars \cite{Dokuch_Quasars} and supermassive black holes within galactic nuclei \cite{Rub_GalNucl,KhlopRubSakh,Dokuch,Grob1}, dark matter \cite{Belotsky,Grob2,PBH,PBH-Carr1,PBH-Dolgov,Gani.IJMPD.2015,Gani.PP.2015}, unidentified sources of gamma-radiation \cite{Belotsky_PBH_gamma_1,Belotsky_PBH_gamma_2}, reionization \cite{PBH-reion1,PBH-reion2,PBH-reion3}, the suppression of intermediate-mass black holes production \cite{Dokuch2,Grob1}.

\section{Acknowledgments}

The authors would like to thank Dr.~K.~M.~Belotsky for useful discussions. The work of A.~A.~K. was supported by Russian Science Foundation, Project \textnumero~15-12-10039. The work of S.~G.~R. was supported by the Ministry of Education and Science of the Russian Federation, Project \textnumero~3.4970.2017/BY. The research fulfilled in the framework of the MEPhI Academic Excellence Project (contract \textnumero~02.a03.21.0005, 27.08.2013) and according to the Russian Government Program of Competitive Growth of Kazan Federal University.

\end{document}